\documentclass[twocolumn, superscriptaddress, prb,floatfix]{revtex4-1}
\usepackage{comment}
\usepackage{color}
\usepackage[usenames,dvipsnames,table]{xcolor}
\usepackage{graphics}
\usepackage{hyperref} 
\usepackage{multirow}
\usepackage{graphicx}
\usepackage{tabularx}
\usepackage{subfigure}
\usepackage[flushleft]{threeparttable}

\usepackage{verbatim}
\usepackage{color}
\expandafter\let\csname equation*\endcsname\relax
\expandafter\let\csname endequation*\endcsname\rela
\usepackage{physics}

\newcommand{\tite}{TiTe$_2$}
\newcommand{\bR}{\textbf{R}}
\newcommand{\bRcal}{\boldsymbol{\Rcal}}
\newcommand{\Rcal}{\mathcal{R}}             
\newcommand{\Dcal}{\mathcal{D}}
\newcommand{\bDcal}{\boldsymbol{\Dcal}}
\newcommand{\bPhi}{\boldsymbol{\Phi}}
\newcommand{\odo}{\overset{0}{\boldsymbol{\mathcal{D}}}}
\newcommand{\od}{\overset{2}{\boldsymbol{\mathcal{D}}}}

\newcommand{\bLambda}{\boldsymbol{\Lambda}}

\newcommand{\bubble}{\scriptscriptstyle{\text{bubble}}}

\newcommand{\tcdw}{T$_{\textrm{CDW}}$}
\newcommand{\te}{T$_e$}

\newcommand{\crystal}{\texttt{CRYSTAL}}

\begin{document}

\title{Theory of the thickness dependence of the charge density wave transition in 1T-TiTe$_2$}

\author{Jianqiang Sky Zhou}
\affiliation{Sorbonne Universit\'e, CNRS, Institut des
Nanosciences de Paris, UMR7588, F-75252, Paris, France}

\author{Raffaello Bianco}
\affiliation{Centro de F\'isica de Materiales (CSIC-UPV/EHU),
             Manuel de Lardizabal pasealekua 5, 20018 Donostia-San 
Sebasti\'an, Basque Country, Spain}

\author{Lorenzo Monacelli}
\affiliation{Dipartimento di Fisica, Universit\`a di Roma La Sapienza, 
Piazzale Aldo Moro 5, I-00185 Roma, Italy}

\author{Ion Errea}
\affiliation{Fisika Aplikatua 1 Saila, Gipuzkoako Ingeniaritza Eskola, University of the Basque Country (UPV/EHU),
Europa Plaza 1, 20018, Donostia San Sebasti\'an, Basque Country, Spain}
\affiliation{Centro de F\'isica de Materiales (CSIC-UPV/EHU), 
    Manuel de Lardizabal pasealekua 5, 20018 Donostia San Sebasti\'an,Basque Country, Spain}
\affiliation{Donostia International Physics Center (DIPC), Manuel de Lardizabal pasealekua 4, 20018 Donostia San Sebasti\'an, Basque Country, Spain}    

\author{Francesco Mauri}
\affiliation{Dipartimento di Fisica, Universit\`a di Roma La Sapienza, 
Piazzale Aldo Moro 5, I-00185 Roma, Italy}
\affiliation{Graphene Labs, Fondazione Istituto Italiano di Tecnologia, Via Morego, I-16163 Genova, Italy}

\author{Matteo Calandra}
\affiliation{Department of Physics, University of Trento, Via Sommarive 14, 38123 Povo, Italy}
\affiliation{Sorbonne Universit\'e, CNRS, Institut des Nanosciences de Paris, UMR7588, F-75252 Paris, France}
\affiliation{Graphene Labs, Fondazione Istituto Italiano di Tecnologia, Via Morego, I-16163 Genova, Italy}

\begin{abstract}
{Most metallic transition metal dichalcogenides undergo charge density wave (CDW) instabilities with similar or identical ordering vectors in bulk and in single layer, albeit with different critical temperatures. Metallic 1T-TiTe$_2$ is a remarkable exception as it shows no evidence of charge density wave formation in bulk, but it displays a stable $2\times2$ reconstruction in single-layer form. The mechanism for this 3D-2D crossover of the transition is still unclear, although strain from the substrate and the exchange interaction have been pointed out as possible formation mechanisms. Here, by performing non-perturbative anharmonic calculations with gradient corrected and hybrid functionals, we explain the thickness behaviour of the transition in 1T-TiTe$_2$. We demonstrate that the occurrence of the CDW in single-layer TiTe$_2$ occurs from the interplay of non-perturbative anharmonicity and an exchange enhancement of the electron-phonon interaction, larger in the single layer than in the bulk. Finally, we study the electronic and structural properties of the single-layer CDW phase and provide a complete description of its electronic structure, phonon dispersion as well as infrared and Raman active phonon modes.  }
\end{abstract}

\maketitle

\section{Introduction}
\label{sec:intro}
Charge density waves (CDWs) are ubiquitous phenomena in condensed matter physics as they appear in many
systems having different electronic structures and dimensionalities. While lots of work has been carried out in one-dimensional (1D) systems with very sharp Fermi surfaces, the mechanism generating charge ordering in higher
dimension is still controversial, mainly because Fermi surfaces are composed of multiple sheets and are not point-like, as in 1D. Furthermore, the variability of the electronic and structural properties substantially affects
the interplay of the three fundamental interactions competing in CDW formation: electron-electron, electron-phonon and anharmonicity.
These factors complicate the explanation of the mechanism responsible for charge ordering and how the latter is affected by external perturbations.

Metallic and semimetallic  transition metal dichalcogenides (TMDs) with chemical formula TX$_2$, where T is a transition metal 
and X is a chalcogene, are among the first 3D systems where 
CDWs were detected and are an example of this large variability as several different ordering vectors and reconstructions
can be found by weakly perturbing the chemical (doping) or structural (e.g., stacking or polytype variation) properties ~\cite{wilsondisalvo}. 
The advent of mechanical exfoliation and the synthesis
of 2D crystals ~\cite{Novoselov10451} added one additional 
parameter, namely the sample thickness, that can be made as thin as that
of a TX$_2$ single layer. It becomes then possible to study
the CDW crossover  
from the bulk to the 2D case. However, up to now, in most of the 
cases the single layer and bulk display similar ordering vectors
and qualitatively similar charge density wave patterns (although the charge density wave critical temperature, \tcdw , can differ in the same compound depending on the thickness of the sample).
The 2H polytypes such as 2H-NbSe$_2$, 2H-TaS$_2$ and 2H-TaSe$_2$, display CDW with the same periodicity in bulk and single-layer form. 
On the contrary, 2H-NbS$_2$ does not show evidence of charge ordering in 
bulk, while contradictory experimental results have been reported in supported single layers, 
as 1H-NbS$_2$ (the notation 1H meaning a single-layer H polytype) on Au(111) does not display a CDW~\cite{NbS2Au111}, 
while 1H-NbS$_2$ on 6H-SiC(0001) endures a $3\times3$ reconstruction~\cite{NbS2CDWmono}. The 1T polytypes display
fairly similar behaviour. 
 1T-TiSe$_2$ undergoes a CDW with the same periodicity both in bulk and single layer, despite some differences in T$_{\rm CDW}$
depending on the substrate or the doping level ~\cite{2DMaterials-2018-substrate-dependent-TC,Wang2018-advmat,Li2016-nature,acs-nano-2017-Duong,Sugawara-ACSNANO-2016,Chen2015-nature-commu,Fang-prb2017}. 1T-TaS$_2$ shows a David star $\sqrt{13}\times\sqrt{13}$ reconstruction both in bulk and in single layer~\cite{Sakabe} (in the bulk the 3D stacking of the stars makes the understanding more complicate and controversial). Finally, 1T-VSe$_2$ reconstructs with a $4\times4$ periodicity both in bulk and in single layer with a non-monotonic dependence of T$_{\rm CDW}$ as a function of layer number~\cite{P_sztor_2017}. 

In this respect, the case of 1T-TiTe$_2$ is definitely surprising and deserves particular scrutiny. Single-layer 1T-TiTe$_2$ displays a $2\times2$ CDW with T$_{\rm CDW}=92$K, but absolutely no CDW occurs  in bulk~\cite{Chen2017-NC}. This result is even more puzzling given the similarity of the electronic structure of 1T-TiTe$_2$ bulk/monolayer with that of 1T-TiSe$_2$ bulk/monolayer, the latter undergoing a CDW with the same ordering vector at all thicknesses.
Harmonic density functional perturbation theory calculations with gradient corrected functionals are unable to explain the main experimental facts, as under this approximation no CDW occurs in TiTe$_2$ neither in bulk nor single layer~\cite{Chen2017-NC}. A recent theoretical work by Guster {\it et al.}~\cite{Guster_2018} claimed that the CDW in single layer could be either due to strain or induced by the exchange interaction. Substrate strain is an unlikely explanation as on the experimental side TiTe$_2$ one-layer thick films are deposed on an incommensurate substrate and the measured lattice parameter is practically the same as in the bulk. Moreover, theory showed that large strains are needed to induce the CDW~\cite{Guster_2018}, a result recently confirmed by strongly epitaxially strained TiTe$_2$ flakes of thickness up to $32$ nm on InAs(111)/Si(111) substrates~\cite{strainedTiTe2}. 

Hartree-Fock exchange is then a plausible explanation, given its importance in bulk and single-layer 1T-TiSe$_2$~\cite{MCalandra-PRL2017,Zhou}. This conclusion is also supported by finite difference harmonic calculations with the HSE06 functional finding the occurrence of the most unstable
phonon mode at the M point of the Brillouin zone (BZ) compatible with a $2\times2$ reconstruction, in agreement with experiments~\cite{Guster_2018}. However, as we will show here, harmonic calculations based on the HSE06 functional predict the occurrence of CDW both in single-layer and bulk 1T-TiTe$_2$, in clear disagreement with experiments. 
Therefore, calculations in literature are unable to explain the reduction of CDW in 1T-TiTe$_2$ as a function of layer thickness.

In this work, we study the vibrational properties of suspended single-layer and bulk 1T-\tite, by accounting for non-perturbative anharmonicity within the stochastic self-consistent
harmonic approximation (SSCHA)~\cite{SSCHA-Ion-prl2013, Raffaello-PRB-2017, Ion-PRB-2014,Monacelli-prb-2018}. The SSCHA is a stochastic variational technique developed by the authors that allows to access the non-perturbative quantum anharmonic free energy and its second derivative 
(i.e., the phonon spectra) from the evaluation of forces on supercells with 
atoms displaced from their equilibrium positions following a suitably chosen Gaussian distribution. The forces can be evaluated by using any
force engine. We show that the interplay between non-perturbative anharmonicity and exchange renormalization of the electron-phonon coupling explains the thickness dependence of CDW in 1T-TiTe$_2$. Moreover, we completely characterize the electronic and vibrational properties of the $2\times2$ reconstruction in single-layer 1T-TiTe$_2$.

\section{Computational details}
\label{sec:methods}

Density-functional theory (DFT) calculations using the Perdew-Burke-Ernzerhof (PBE)~\cite{PRL-1996-PBE} and HSE06~\cite{HSE2003,HSE2006} exchange-correlation functionals are carried out using the \texttt{CRYSTAL}~\cite{crystal-17} and the \textsc{Quantum-ESPRESSO}~\cite{QE-2009,QE-2017} packages. The optimized lattice parameters for the undistorted CdI$_2$ phase of bulk and single-layer \tite\ can be found in Tab. \ref{tab:table-lattice-atp}. As it can be seen, both HSE06 and PBE accurately describe the in-plane lattice parameter but substantially overestimate the interlayer distance in the bulk, due to the lack of Van der Waals forces. A practical and common way to avoid this problem in TMDs is to adopt the experimental measured lattice parameters $a=3.777$ \AA\ and for the bulk, with $c=6.495$ \AA~\cite{Chen2017-NC}. In the single-layer case, we used a 12.99 \AA\ vacuum region to avoid interactions between periodic images. We perform geometrical optimization of internal coordinates.  For the \texttt{CRYSTAL} code we use an all-electron molecular def2-TZVP basis set~\cite{weigend2005} reoptimized for solid state calculations~\cite{MCalandra-PRL2017} for the Ti atom and a pVDZ-PP basis set for the Te atom~\cite{Te-basis-1,Te-basis-2}. For \textsc{Quantum-ESPRESSO} HSE06 fully-relativistic calculations, we used norm-conserving ONCV pseudopotentials from the Pseudo-dojo library~\cite{pseudodojo} (high accuracy). The electronic and harmonic phonon bands are calculated using the same $\Gamma$ centered ${\bf k}$-points mesh and electronic temperature (i.e., smearing in Fermi-Dirac function, see Tab. \ref{tab:table-ksm} in Appendix A for more technical details). The ${\bf k}$-points mesh is rescaled according to the size of supercells (e.g., a $36\times36\times1$ ${\bf k}$-points mesh in $1\times1\times1$ cell becomes $9\times9\times1$ in a $4\times4\times1$ cell ). Since the harmonic phonon frequency is very sensitive to the chosen ${\bf k}$-points sampling and electronic temperature \te , the convergence of the lowest phonon frequency with respect to the ${\bf k}$-points sampling and electronic temperature is carefully investigated for both DFT functionals (see Fig. \ref{fig:convergence-highT} in Appendix A). The HSE06 forces needed for the stochastic self-consistent harmonic approximation (SSCHA)~\cite{SSCHA-Ion-prl2013, Raffaello-PRB-2017, Ion-PRB-2014,Monacelli-prb-2018} are computed with the \crystal\ code. 

\begin{table}[ht!]
\caption{\label{tab:table-lattice-atp}
Completely optimized structural parameters for bulk and single-layer \tite\ in the undistorted CdI$_2$ phase
with space group P$\bar{3}$m1 (number 164) compared to experiments. The quantity $z_{\rm Te}$ is the only internal parameter not determined by symmetry, namely the tellurium distance from the plane of Ti atoms. The $z_{\rm Te}$ values in brackets are obtained assuming the experimental lattice parameters $a=3.777$\AA~  and  $c=6.495$\AA~  and optimizing only internal coordinates. We use the experimental lattice parameters and the values of $z_{\rm Te}$ in brackets in all  bands and phonon calculations. }
\begin{center}
\begin{threeparttable}
\begin{footnotesize}
\begin{tabular}{c|c|c|c|c}
 \hline
 \multicolumn{5}{c}{bulk} \\
 \hline\hline
 & PBE & HSE06 & EXP ~\cite{Chen2017-NC} & EXP ~\cite{Patel1985}\\
 $a$ & 3.77&3.79&3.777& 3.768\\
 $c$ & 6.94 & 6.94&6.495& 6.524\\
 $z_{\rm Te}$ & 1.73(1.71)&1.69(1.68)& &1.66 \\
 \hline
  \multicolumn{5}{c}{monolayer} \\
 \hline\hline
 $a$ & 3.765 &3.788& 3.78 & \\
 $z_{\rm Te}$ & 1.74(1.73)& 1.69(1.70)& &  \\
 \hline
\end{tabular}
\end{footnotesize}
\end{threeparttable}
\end{center}
\end{table}

\section{Results and Discussion}


\subsection{Electronic structure of the undistorted \texorpdfstring{CdI\textsubscript{2}}{} phase}
\begin{figure*}[ht!]
\centering
   \includegraphics[width=0.95\textwidth]{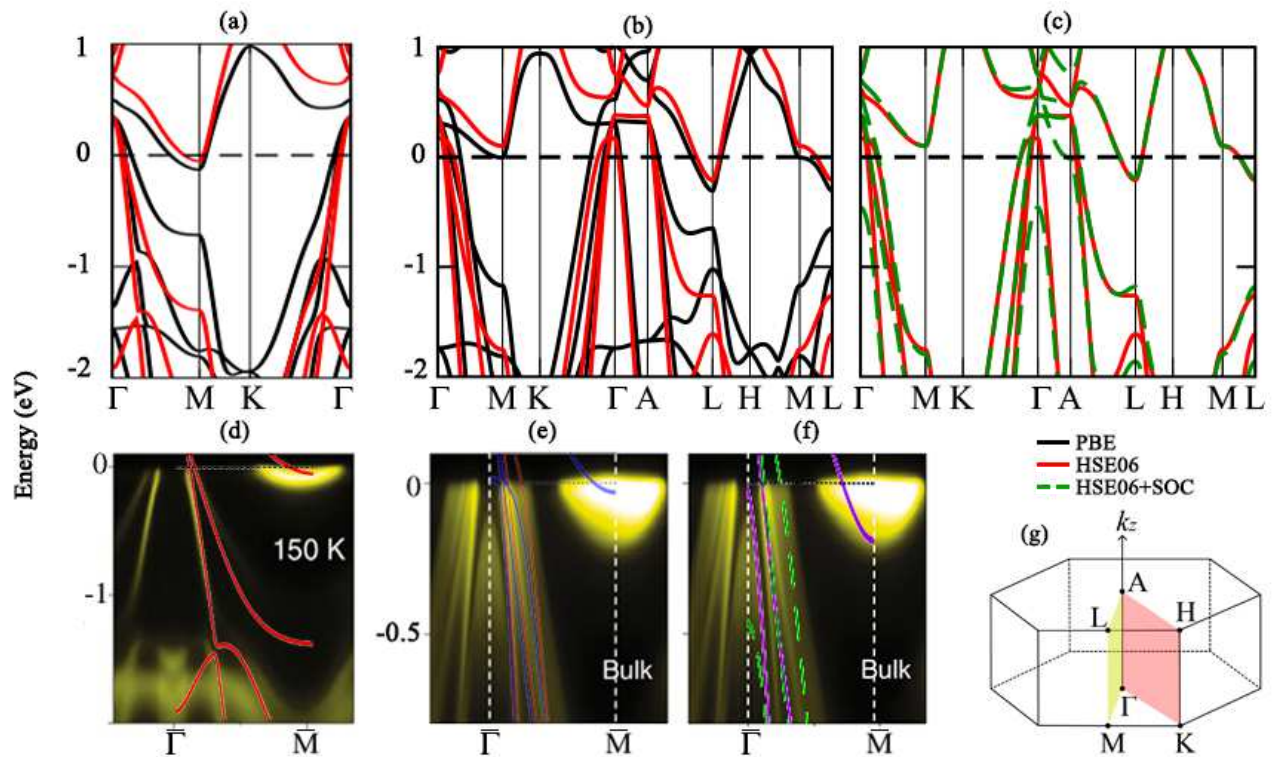}
    \caption{Calculated monolayer (a) and bulk (panels (b) and (c)) electronic band structures using the PBE (black),  HES06 (red), and HSE06+SOC (green) functionals. In  panel (b) and (c) the band dispersion along the ML direction illustrates strong the $k_z$ dependence. In panel (d-f) we present the comparison between calculated and measured ARPES bands where (d) represents monolayer ARPES VS HSE06 along $\Gamma$M, (e) shows the bulk ARPES VS HSE06 along $\Gamma$M (in red) and along AL (in blue), (f) shows the bulk ARPES VS HSE06+SOC along $\Gamma$M (in green) and along AL (in violet). Panel (g) shows the BZ of the 1T-\tite\ and the high symmetry points. The HSE06+SOC calculation is performed with \textsc{Quantum-ESPRESSO}, all other calculations with \crystal.}
    \label{fig:electronic-bands}
\end{figure*}


\subsubsection{Theory.~}

The electronic structure of the undistorted CdI$_2$ phase calculated
from two DFT functionals, the scalar relativistic semilocal PBE and the scalar relativistic hybrid HSE06, is shown
in Fig.~\ref{fig:electronic-bands} (a) and (b) for the monolayer
and bulk, respectively. Both approximations give a semimetallic
ground state, in agreement with several previous first-principles band structure calculations~\cite{Guster_2018,PhysRevB.54.2453,PhysRevB.29.6797,PhysRevB.97.085107}. The semimetallic character is due to the band overlap
between the Te hole pockets at zone center (multiband in nature) and the Ti $3d$ electron-pocketcat M in the monolayer and L in the bulk. In the bulk the band-overlap between L and $\Gamma$ is larger than for the single-layer case between M and $\Gamma$. The inclusion of screened exchange within the HSE06 functional has two main effects: (i) to reduce the band-overlap between zone center and M (L) in the single layer (bulk) and (ii) to substantially increase the Fermi velocity of the Te bands close to zone center.

The inclusion of relativistic effects has minor consequences for the
single layer, as shown in appendix B for the PBE semilocal functional~\cite{suppmat}. In the bulk, it affects mainly the bands at zone center where there are three
Te bands forming three hole pockets. In the absence of spin-orbit coupling (SOC), two of these bands (arising from Te $p$ orbitals) are degenerate at $\Gamma$, while the third one is not. SOC splits the two Te degenerate bands at zone
center, downshifting one of the two and upshifting the other, as shown in Fig.~\ref{fig:electronic-bands} (c) for the HSE06 case
and in appendix B for the PBE case, respectively.
However, while in PBE the electron-electron interaction still leaves a portion of the lower of the two bands unoccupied, the combined effect of HSE06 and SOC leads to a completely occupied Te band at zone center (see Fig.~\ref{fig:electronic-bands} (c)). As we will show in the next paragraph, the combined effect of exchange and relativistic effects is needed to solve a
long standing controversy in ARPES spectra.

\subsubsection{Comparison with ARPES.~} 

ARPES experiments show a semimetallic nature for both single-layer and bulk \tite~\cite{Chen2017-NC,PhysRevB.54.2453,PhysRevB.74.195125} in its CdI$_2$ undistorted phase.
The comparison between the calculated electronic structures and ARPES data for the single layer and bulk are shown in Fig.~\ref{fig:electronic-bands} (d,e,f).
The HSE06 electronic structure and ARPES data are in excellent
agreement for the single-layer case (panel d). The PBE approximation
gives unrealistically too low Fermi velocities for the Te band close to zone center forming the largest hole pocket and a too large occupation of the Ti $3d$ pocket at M. 

In the bulk case, ARPES spectra of TiTe$_2$ have been measured in 
several works~\cite{Chen2017-NC,PhysRevB.54.2453,PhysRevB.74.195125}, we compare here with the most recent work of~\cite{Chen2017-NC}.
Some care is needed in comparing theory and experiments in bulk
due to the very strong $k_z$ band dispersion and the fact that $k_z$ is not a good quantum number in ARPES. It is then not obvious that measurements really probe the bands at $k_z=0$. This issue
has been carefully addressed in Ref.~\cite{PhysRevB.54.2453,PhysRevB.74.195125}, where the magnitude of $k_z$ dispersion along ML was estimated to be included between $20$ and $100$ meV binding energy. Experimentally, the determination of $k_z$ is complicated by the presence
of what are usually labeled as {\it non-free electrons final state effects} (i.e., the approximation that the final-state surface-perpendicular dispersion $\epsilon_{\bf k_{\perp}}$ is assumed to be parabolic breaks down)~\cite{PhysRevB.74.195125}.
For this reason theoretical description of ARPES for bulk TiTe$_2$ is challenging.

In Fig.~\ref{fig:electronic-bands} panels (e,f) we superimpose the bands along AL and along $\Gamma$M (plotted with different colors) to the experimental ARPES data. At the $\bar{\rm M}$ point, ARPES would be consistent with the calculated electronic structure at a $k_z$ somewhere close to half the distance
between $\Gamma$ and M. 

At zone center, four or five bands are measured in ARPES, as it can be seen in Fig.~\ref{fig:electronic-bands} (d) or in Fig.~3 of Ref.~\cite{Chen2017-NC} panel (a). However, only three bands occur close to $\Gamma$ in the calculation, thus part of these bands are necessarily related to other values of $k_z$. As it can be seen from Fig.~\ref{fig:electronic-bands} (d,e), this is consistent with the calculated band structure and its $k_z$ dispersion that generates more shadow bands. The non-relativistic HSE06 bands (panel d) are in better
agreement with experiments if contributions from scattering at $k_z\ne0$ are assumed to occur. Still, if relativistic effects are neglected, there is one important
disagreement between theory and experiments, namely the fact
that ARPES data show the presence of a completely occupied parabolic band at $\approx-0.135$ eV binding energy (see experimental data in Fig.~\ref{fig:electronic-bands} (d,e))  that is missing in all samples below three layers (see Fig.~3 of Ref.~\cite{Chen2017-NC} panel (a)) and in all PBE calculations (including or neglecting SOC) for any value of $k_z$.
This band was detected as a very broad feature in previous experimental ARPES work~\cite{PhysRevB.54.2453} (hatched region in Figs.~5, 13, and 14), however, its origin is unclear in literature. It was proposed to be due to many body effects beyond the single particle approximation.

Here we demonstrate, on the contrary, that this band arises from the combined effect of screened exchange and relativistic effects,
as shown in Fig.~\ref{fig:electronic-bands}, panel (f), as discussed in the previous subsection.
In the HSE06 relativistic calculation the band is somewhat lower in energy at zone center than in experiments. This is most likely due to the $n$-doping occurring in TiTe$_2$ samples. We also stress that the exact position of this band is extremely sensitive on the Te distance from the plane and differences of $0.01$ \AA\ leads to a sizeable energy shift. Thus our calculation explains this feature without invoking any many body effects and solves the long standing ARPES controversy.


\subsection{Harmonic phonon dispersion for the \texorpdfstring{CdI\textsubscript{2}}{} undistorted phase}

\begin{figure}[ht!]
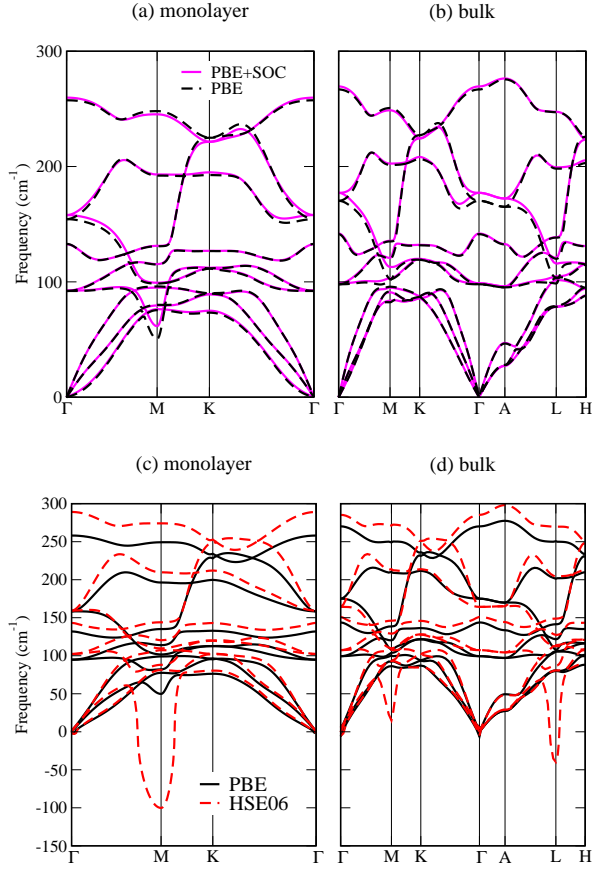

    \centering
    \includegraphics[width=0.9\linewidth]{harmonic-phonon-monolayer-bulk-pbe-soc.eps}\vspace{0.5cm}
    \includegraphics[width=0.9\linewidth]{harmonic-phonon-mono-bulk-pbe-hse06.eps}
    \caption{Top panels: the monolayer and bulk harmonic phonon dispersion in PBE with and without spin orbit coupling calculated with the QUANTUM-ESPRESSO code. Bottom panels: PBE and HES06 phonon dispersion neglecting spin-orbit coupling calculated with the \crystal\ code.}
    \label{fig:harmonic-phonon-highT}
\end{figure}

The harmonic phonon dispersion for bulk and single layer using the PBE and HSE06 functionals with and without SOC are shown in Fig.~\ref{fig:harmonic-phonon-highT}. The PBE phonon dispersion neglecting SOC are found to be in good agreement with previous calculations of Refs.~\cite{Chen2017-NC,Guster_2018} and show only positive phonon frequencies and no CDW formation, neither in bulk nor in monolayer. The inclusion of SOC leads to negligible differences both in bulk and monolayer. This means that the changes in the electronic structure due to the relativistic effects have only marginal consequences for the CDW formation.

The single-layer and bulk PBE phonon dispersion are, however, markedly different as in the former case a softening occurs at the M point, while no softening is seen in the bulk at the L point. The result changes completely if the HSE06 functional is used, as now both the bulk and the single layer show imaginary phonon frequencies (depicted as negative) at the L and M points, respectively.
Thus, within HSE06 and at the harmonic level, {\it both} single layer and bulk do display a CDW, in disagreement with experimental data showing absence of charge ordering in the bulk.
Therefore, the exchange interaction alone is not sufficient to explain the thickness dependence of the CDW in \tite, as previously proposed~\cite{Guster_2018}. 
In the case of the single layer, we also calculated the energy gain by the distortion in a $2\times2$ supercell by displacing the atoms along the phonon pattern of the structural distortion, finding it to be approximately $3.13$ meV per Ti atom,
a value in excellent agreement with the HSE06 plane waves calculation carried out in Ref.~\cite{Guster_2018}. 

The different behaviour of the harmonic phonon dispersion at the L (bulk) and M (single layer) points for the different functionals can be due to two effects: (i) differences in the electronic structure and (ii) differences in the electron-phonon matrix elements. The softening is indeed due to the real part of the electron-phonon self-energy phonon~\cite{CalProf}:
\begin{equation}
\Pi^{\prime}_{\bf \nu}({\bf q})={\cal P}\frac{1}{N_k}\sum_{{\bf k},nm}\frac{|g_{{\bf k}n,{\bf k}+{\bf q}m}^{\nu}|^2\left(f_{{\bf k}n}-f_{{\bf k}+{\bf q}m}\right)}{\epsilon_{{\bf k}n}-\epsilon_{{\bf k}+{\bf q}m}}     
\end{equation}
where ${\cal P}$ label the principal part, $N_k$ the number of {\bf k} points used in the calculations ($120\times120$ and $80\times80\times20$ for the single layer and bulk, respectively), $\epsilon_{{\bf k} n}$ are the band energies
and $f_{{\bf k} n}$ the related Fermi functions.
Finally, $g_{{\bf k}n,{\bf k}+{\bf q}m}^{\nu}$ is the electron-phonon matrix element for the mode $\nu$.
The softening is then obtained as $\omega_{{\bf q}\nu}^2=\Omega_{{\bf q}\nu}^2+2\Omega_{{\bf q}\nu}\Pi^{\prime}_{\nu}({\bf q})$.

In order to understand the mechanism responsible for the softening, we use maximally localized Wannier functions~\cite{MostofiWannier,MVcomposite,MVentangled} and calculate the real part of the bare susceptibility with constant matrix elements, namely
\begin{equation}
\chi_0({\bf q})=
{\cal P}\frac{1}{N_k}\sum_{{\bf k},nm}
\frac{f_{{\bf k}n}-f_{{\bf k}+{\bf q}m}}{\epsilon_{{\bf k}n}-\epsilon_{{\bf k}+{\bf q}m}}     
\end{equation}
The quantity $\chi_0$ is related to $\Pi^{\prime}_{\bf \nu}({\bf q})$ in the approximation of constant electron-phonon matrix elements, i.e., $g_{{\bf k}n,{\bf k}+{\bf q}m}^{\nu}=g$, as $\Pi^{\prime}_{\bf \nu}({\bf q})=g\chi_0({\bf q})$. Thus, it probes the effect of the
electronic structure on the softening, but not those related to the dependence of the matrix element $g_{{\bf k}n,{\bf k}+{\bf q}m}^{\nu}$ on the exchange correlation functional or on $\bf k$ and band index.
\begin{figure}[ht!]
\centering
   \includegraphics[width=0.9\columnwidth]{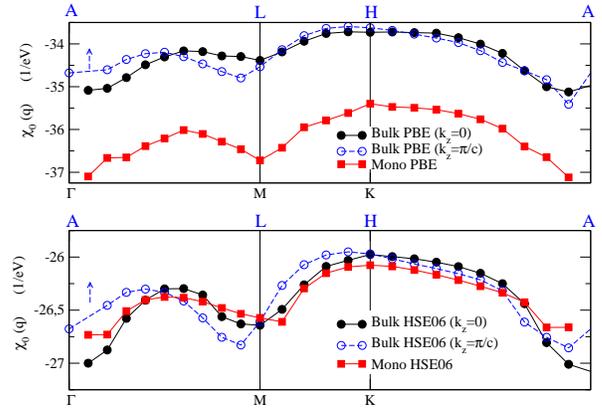}
    \caption{Calculated real part of $\chi_0({\bf q})$ for bulk and monolayer TiTe$_2$ using PBE and HSE06.}
    \label{fig:chi0}
\end{figure}
As it can be seen in Fig.~\ref{fig:chi0}, the ${\bf q}$ dependence of $\chi_0({\bf q})$ is very similar in the different cases (bulk and single layer), meaning that the main effect of the exchange functional on the harmonic spectra is due to a renormalization of the electron-phonon matrix elements and not of to a change in the electronic structure.

\subsection{Anharmonicity and charge ordering}
\begin{figure}[ht!]
    \centering
    \includegraphics[width=\linewidth]{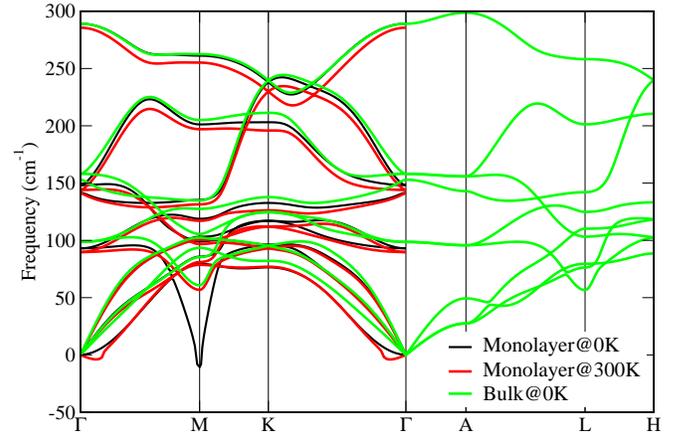}
    \caption{The anharmonic phonon dispersion of bulk (at 0K) and monolayer (at 0K and 300K) using the HSE06 functional from the \crystal\ code.}
    \label{fig:D3V-mono-0kand300k-bulk-0k}
\end{figure}
The calculations performed up to now neglect phonon-phonon scattering (anharmonicity) and its tendency to stabilize the lattice.
In what follows, we carry out non-perturbative anharmonic calculations within the stochastic self-consistent harmonic approximation 
using HSE06 as the force engine. In particular, we evaluate the temperature dependent dynamical matrix 
\begin{equation}
\bDcal =  \boldsymbol{M}^{-\frac{1}{2}}\eval{\pdv{F}{\bRcal}{\bRcal}}_{\bRcal_{eq}} \boldsymbol{M}^{-\frac{1}{2}}
\label{eq:Hessian}
\end{equation}
where $\boldsymbol{M}$ is the matrix of the ionic masses $M_a$ with $M_{ab}=\delta_{ab}M_a$ and ${\bf {\cal R}}$ are the coordinates of the centroids (i.e., average value of the atomic positions over the ionic wavefunction). 
The free energy and its second derivative can be obtained by performing appropriate stochastic averages over the atomic forces on
supercells with ionic configurations obtained by displacing the atoms randomly from the equilibrium position and following a Gaussian distribution~\cite{SSCHA-Ion-prl2013, Ion-PRB-2014,Raffaello-PRB-2017}.
While the free energy converges fairly quickly with the number of ionic configurations, the Hessian of the free energy is more noisy and
a larger number of samples to converge. We use from $600-900$ and $400$ force calculations for the single layer and bulk, respectively.

As the all-electron HSE06 force engine is computationally very expensive, we perform the calculation on a $4\times4$ supercell in the monolayer (i.e., 48 atoms containing the wavevector ${\bf q}={\bf M}$) and a $4\times4\times2$ supercell in the bulk case (i.e., 96 atoms containing the wavevector ${\bf  q}={\bf L}$). We know from previous studies~\cite{Zhou} that this supercell is large enough to describe charge density wave formation in the similar compound 1T-TiSe$_2$, but T$_{\rm CDW}$ comes out somewhat underestimated (accurate determination of T$_{\rm CDW}$ requires larger supercells, unaffordable with the HSE06 functional as the force engine). 
For this reason we perform calculations at zero and room temperature for the monolayer and at zero temperature for the bulk. We then determine if anharmonicity can stabilize or not the lattice. We expect that on larger supercells, the instability at the M point in the monolayer will be slightly stronger.
More details on the anharmonic calculation and the magnitude of the different terms occurring in Eq.~\ref{eq:Hessian} are given in appendix C.

The results of the anharmonic calculation for the high-$T$ CdI$_2$ phase are shown in Fig.~\ref{fig:D3V-mono-0kand300k-bulk-0k}. As it can be seen, in both cases anharmonicity tends to stabilize the lattice and at $T=0$K the
(positive) anharmonic correction to the soft mode at the M point for the monolayer  is similar in magnitude, to the corresponding one at the L point for the bulk. 
However, as the harmonic frequency is substantially much softer in the monolayer case, the anharmonic phonon frequency at the M point remains imaginary in the monolayer case, consistent with the experimental finding of a $2\times2$ reconstruction. On the contrary, in the bulk case, anharmonicity completely removes
the imaginary phonon frequency at the L point, stabilizing the lattice and removing the charge ordering found at the harmonic level, again in agreement with the
experimental findings~\cite{Guster_2018}. At $T=300$K, the monolayer displays stable phonon frequencies.
Thus, the thickness evolution of the CDW in TiTe$_2$ is due to a competition of two effects, anharmonicity and the electron-phonon interaction.
While anharmonicity has similar magnitude in bulk and single layer, the electron-phonon interaction leads to much more unstable harmonic phonons in the single layer at the M
point than in the bulk at the L point. At the PBE level, however, the electron-phonon correction to the phonon
frequency is not large enough to induce charge ordering. 
The HSE06 is responsible for a stronger  electron-phonon interaction than in the PBE case. The HSE06 harmonic phonon dispersion displays CDWs both in bulk and in single layer, in disagreement with experiments. However, the single layer harmonic phonon frequencies are substantially more unstable than the ones in the bulk. Anharmonicity, similar in magnitude for the two case, removes the CDW in the bulk but not in the single layer, in perfect agreement with experiments.

\subsection{Structural, electronic and vibrational properties of the single-layer charge ordered phase}

After explaining the appearance of CDW in single-layer TiTe$_2$, we study the low temperature
$2\times2$ phase. Structural data for the single layer are given in Tab.~\ref{tab:atp-lowT} obtained from geometrical optimization of forces and neglecting quantum effects. The distortion is analogous to that found in 
a single-layer TiSe$_2$. 

\begin{table}[ht!]
\begin{center}
\begin{footnotesize}
\begin{tabular}{ l c c c }
\textrm{atoms}&
x & y & z \\
\hline\hline 
Ti & 0.0 & 0.0  & 0.0 \\
Te & -0.33311 & -0.16931&  0.13099 \\
Ti&    0.49038  &  0.0  & 0.0 \\
Te&   1/3          & -1/3&      -0.13034 \\
\hline\hline 
\end{tabular}
\end{footnotesize}
\end{center}
\caption{\label{tab:atp-lowT}
 Distorted structure of single-layer 1T-TiTe$_2$ within HSE06. The space group is $P321$ (number $150$), as we use 3D labeling of space groups assuming an infinite distance between the TiTe$_2$ layers. The Wyckoff positions are given as components with respect to the conventional cell. The in-plane lattice parameter is twice the experimental one
 of the CdI$_2$ phase.}
\end{table}

\begin{figure}
    \centering
    \includegraphics[width=0.9\linewidth]{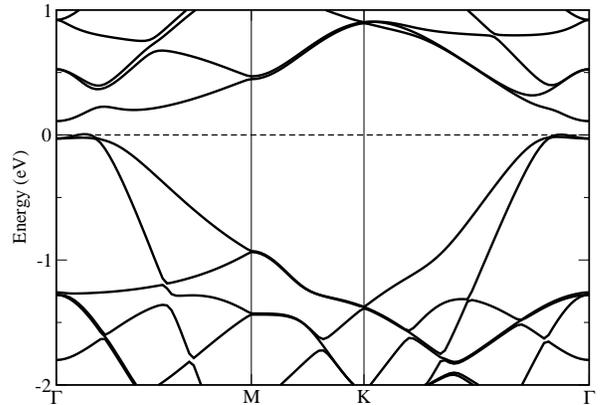}
    \caption{The HSE06 electronic band structure of monolayer \tite\ in its low temperature phase ($2\times2$ superstructure). }.
    \label{fig:electronic-bands-monolayer-tite2-lowT}
\end{figure}

The HSE06 electronic structure of the distorted phase is shown in Fig.~\ref{fig:electronic-bands-monolayer-tite2-lowT}. An indirect gap of $0.1395$ eV is found to occur at the Fermi level. In STM experiments a pseudogap of $\approx0.028$ eV at 42K~\cite{Chen2017-NC} is found in the CDW phase, substantially smaller. A similar overestimation of the gap in the low-$T$ phase by the HSE06 functional is found in single-layer TiSe$_2$.
Both these overestimations could  be due to the neglect of nuclear quantum fluctuations in the low-$T$ phase and their consequences on the electronic gap.  

\begin{figure}
    \centering
    \includegraphics[width=\linewidth]{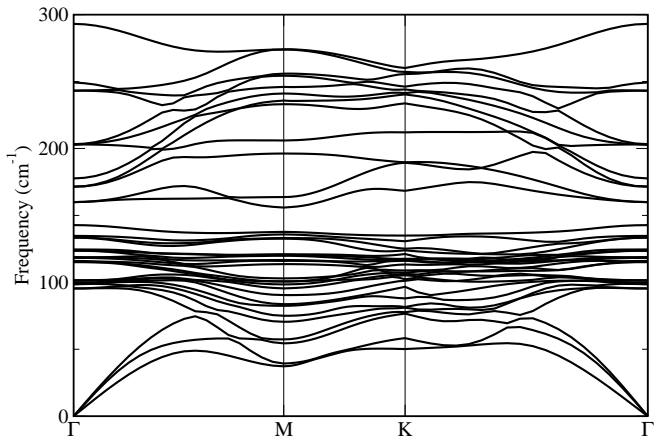}
    \caption{The HSE06 harmonic phonon dispersion of monolayer \tite\ in its low temperature phase ($2\times2$ superstructure). }
    \label{fig:harmonic-phonon-monolayer-tite2-lowT}
\end{figure}

In order to test the stability of the low-$T$ phase, we calculate the harmonic phonon dispersion using the HSE06 functional. The results are shown in Fig.~\ref{fig:harmonic-phonon-monolayer-tite2-lowT}. We find dynamically stable phonon frequencies.  
As the Raman and infrared phonon frequencies can in future be used to determine the structural properties of the monolayer, 
we report in detail the zone center Raman and infrared modes in Tabs.~\ref{tab:Raman} and~\ref{tab:IR}, respectively. Furthermore we report their decomposition in terms of the harmonic phonon modes
of the undistorted structure, along the lines of what have been done in Ref.~\cite{MCalandra-PRL2017} (see supplemental materials in Ref.~\cite{MCalandra-PRL2017} for more details). 

\begin{table*}[ht!]
\caption{\label{tab:IR}%
Infrared active modes (cm$^{-1}$) in the CDW phase of TiTe$_2$ single-layer and their decomposition in harmonic undistorted phonon modes. The point(s) \textbf{q} in the Brillouin zone and the frequency(s) $\omega_{undist.}$ label the frequencies and the phonon-momenta in the  undistorted cell having the largest overlap with the zone phonon modes of the distorted cell. The number in parenthesis is the overlap in percentage (only those are bigger than 10\% are shown). }
\begin{center}
\begin{footnotesize}
\begin{tabular}{ l c c c }
Symmetry& HES06 (CDW phase) & \textbf{q} 
&$\omega_{undist.}$ \\
\hline\hline
E & 79.46 & M & 82.52(50) + -102.09(25) + 88.70(17)  \\
A2 & 81.89 & M & 82.96(97) \\
E & 83.27 & M & 82.52(38) + 88.70(23) + -102.09(18)  \\
E & 91.79 & M + $\Gamma$ + M & 88.70(39) + 101.17(25) + -102.09(15) \\
E & 104.09 & $\Gamma$ + M & 101.2(29) + 108.1(21) + -102.09(14) + 108.87(14) \\
A2 & 105.54 & M & 108.99(81)\\
A2 & 106.84 &M & 108(67)+108.1(21) \\
E & 107.39 & M & 108.89(59)+108.87(17)+108.1(10)\\
E & 109.55 & $\Gamma$+3M & 101.17(31)+108.08(27)+-102.09(15)+108.1(13)\\
E & 120.21& M & 122.66(85) \\
E &142.63 & $\Gamma$ & 143.8(92)\\
E &172.79 & $\Gamma$ & 158.3(89) \\
A2 & 207.72 & M & 208.9(99)\\
E & 212.28& M & 208.87(97) \\
A2 & 276.54 & M & 275.18(100) \\
E & 278.97 & M & 274.97(99)\\
A2 & 293.53 & $\Gamma$ & 290.04(99)\\
\hline\hline
\end{tabular}
\end{footnotesize}
\end{center}
\end{table*}

\begin{table*}[ht!]
\caption{\label{tab:Raman}%
Raman active modes (cm$^{-1}$) identified in the CDW phase of TiTe$_2$ monolayer. The point(s) \textbf{q} in the Brillouin zone and the frequency(s) $\omega_{undist.}$ of the mode(s) of undistorted cell overlapping with the modes in the distorted cell are also reported, with the overlap percentage (only those are bigger than 10\% are shown) between parentheses. }
\begin{center}
\begin{footnotesize}
\begin{tabular}{ l c c c }
Symmetry& HES06 (CDW phase) & \textbf{q} 
&$\omega_{undist.}$ \\
\hline\hline
E & 79.46 & M &82.52(50) + -102.09(25) + 88.70(17)  \\
E & 83.27 & M & 82.52(38) + 88.70(23) + -102.09(18)  \\
A1 & 85.77 &M & 88.64(94)\\
E & 91.79 & M + $\Gamma$ + M & 88.70(39) + 101.17(25) + -102.09(15) \\
E & 104.09 & $\Gamma$ + M & 101.2(29) + 108.1(21) + -102.09(14) + 108.87(14) \\
E & 107.39 & M & 108.89(59)+108.87(17)+108.1(10)\\
E & 109.55 & $\Gamma$+M & 101.17(31)+108.08(27)+-102.09(15)+108.1(13)\\
A1 & 111.97 & M+$\Gamma$ & 122.91(43)+ -100.91(40)+145.96(11)\\
E & 120.21& M & 122.66(85) \\

A1 &130.72 & M+$\Gamma$+M &  122.91(52)+145.96(27)+-100.91(18) \\
E &142.63 & $\Gamma$ & 143.8(92)\\
A1& 143.33 & $\Gamma$ & 144.71(89)+145.96(10) \\
A1 &157.03 & $\Gamma$+M & 145.96(51)+-100.91(38) \\
E &172.79 & $\Gamma$ & 158.3 (89) \\
E & 212.28& M & 208.87(96) \\
E & 278.97 & M & 274.97(99)\\
\hline\hline
\end{tabular}
\end{footnotesize}
\end{center}
\end{table*}

\section{Conclusions}
\label{sec:conclusion}

In this work we studied the 2D-3D crossover of the CDW transition in metallic 1T-TiTe$_2$. This system
is a remarkable exception between dichalcogenides as it shows no evidence of CDW formation in bulk, but it displays a stable $2\times2$ reconstruction in single-layer form (most of metallic dichalcogenides display similar reconstructions in both bulk and single-layer form). In literature, the mechanism of the transition is  unclear. Strain from the substrate and the exchange interaction have been pointed out as possible formation mechanisms. By
performing non-perturbative anharmonic calculations with gradient corrected and hybrid functionals, we explained the thickness behaviour of the transition 1T-TiTe$_2$. 
We first showed that, at the harmonic level, semilocal functionals fail in describing the CDW transition occurring in the monolayer, while the HSE06 functional predicts the occurrence of a CDW both in bulk and single layer, in disagreement with experiments. At the harmonic level,
the presence of CDW at all thicknesses within HSE06 is not due to a change in the electronic structure but mostly to an exchange renormalization of the electron-phonon matrix element.

Our non-perturbative anharmonic calculations show that the occurrence of CDW in single-layer TiTe$_2$ comes from the interplay of non-perturbative anharmonicity and an exchange enhancement of the electron-phonon interaction, leading to more unstable harmonic phonon modes in the single layer than in the bulk.  Indeed, anharmonicity tends to stabilize both structure in a similar way, however, the larger instability present in the single layer at the harmonic level is not completely removed, while it is totally suppressed in the bulk.

Finally, in an effort to better identify the properties of the single-layer 1T-TiTe$_2$ $2\times2$ CDW phase, yet not fully characterized experimentally, we study its electronic and structural properties and  we provide a complete description of infrared and Raman active phonon modes in terms of the backfolding of the vibrational modes from the undistorted structure. 

\section*{Acknowledgements}
{Computational resources were granted by PRACE (Project No. 2017174186) and from
IDRIS, CINES and TGCC (Grant eDARI 91202 and Grand Challenge Jean Zay). M.C., F. M. , J.S.Z. and L.M. acknowledge support from the Graphene Flaghisp core 2 (Grant No. 785219). M.C. and J.S.Z. acknowledge support from Agence nationale de la recherche (Grant No. ANR-19-CE24-0028). F. M. and L. M. acknowledge support by the MIUR PRIN-2017 program, project number 2017Z8TS5B. }

\section*{Appendix A: Convergence tests}

\begin{figure}
    \centering
    \includegraphics[width=0.9\linewidth]{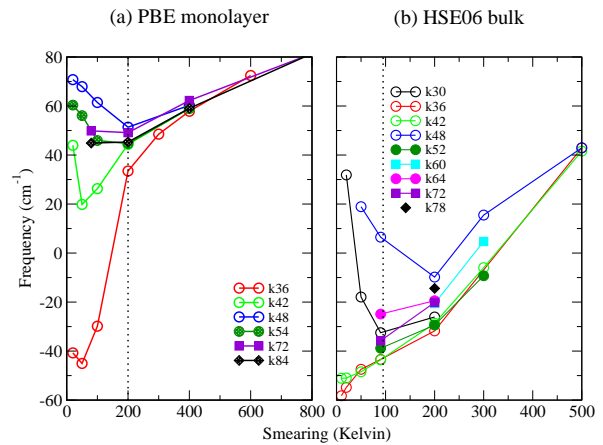}
    \caption{The convergence studies on the monolayer \tite\ using PBE in its normal phase (a) and bulk \tite\ using HSE06 (b). The vertical dotted line represents the converged smearing temperature. The legend shows different ${\bf k}$-points mesh, e.g., k36 in (a) and (b) represents a $36\times36\times1$ mesh and $36\times36\times8$, respectively. Note that in (b) we show the in-plane mesh convergence, and finally we converge the $k_z$ direction as concluded in Tab. \ref{tab:table-ksm} .}
    \label{fig:convergence-highT}
\end{figure}

In the main text, we have shown harmonic phonon calculations using well converged  ${\bf k}$-points mesh and Fermi-Dirac smearing that are summarized in Tab.~\ref{tab:table-ksm}. Here in Fig.~\ref{fig:convergence-highT} we show the  convergence of the lowest energy (unstable) phonon mode at M in the undistorted monolayer and at L using different functionals.
In the table we list the converged parameters for all vibrational calculations.
\begin{table}[ht!]
\caption{\label{tab:table-ksm}%
Converged parameters for the harmonic phonon calculations.
}
\begin{center}
\begin{footnotesize}
\begin{tabular}{ l c c}
\textrm{methods}&
\textrm{${\bf k}$-points}
&\textrm{T$_e$ (Kelvin)} \\
\hline\hline
PBE (bulk)& $48\times 48 \times12$ & 315 \\
HSE06 (bulk)& $36\times36\times12$ & 95 \\
PBE (mono-HT) & $36\times36\times1$ & 200  \\
HSE06 (mono-HT) & $48\times48\times1$ & 32 \\
HSE06 (mono-LT) & $36\times36\times1$ & 30 \\
\hline\hline
\end{tabular}
\end{footnotesize}
\end{center}
\end{table}

\section*{Appendix B: Relativistic effects and semilocal functionals.}

\begin{figure}
    \centering
    \includegraphics[width=0.9\linewidth]{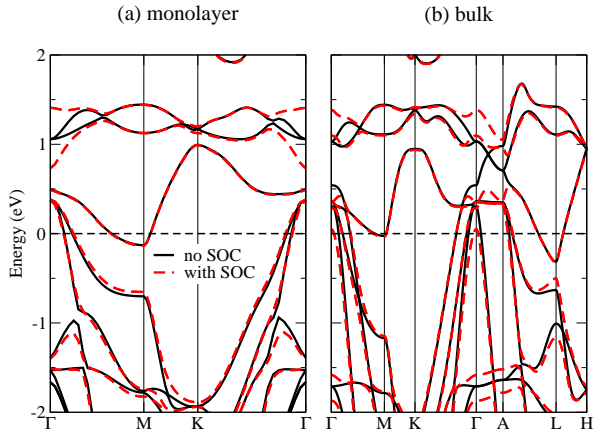}
    \caption{The effects of SOC in monolayer and bulk \tite\ electronic band structures using PBE. Note the absence of a completely filled Te band at zone center (in disagreement with experiments) and in the HSE06 relativistic calculation in Fig. \ref{fig:electronic-bands}}.
    \label{fig:socmono}
\end{figure}

We show in Fig.~\ref{fig:socmono} the effect of SOC on the gradient corrected electronic structure of bulk TiTe$_2$. As it can be seen SOC is completely negligible in the monolayer and has slightly larger consequences in the bulk. In the bulk, however, SOC on top of gradient corrections fail in reproducing the occurrence of a completely filled Te band at zone center (see Fig.~\ref{fig:electronic-bands} panels (d,e,f)). This failure is corrected by HSE06+SOC.

\section*{Appendix C: different contributions to the free energy Hessian}

\begin{figure}
    \centering
    \includegraphics[width=0.9\linewidth]{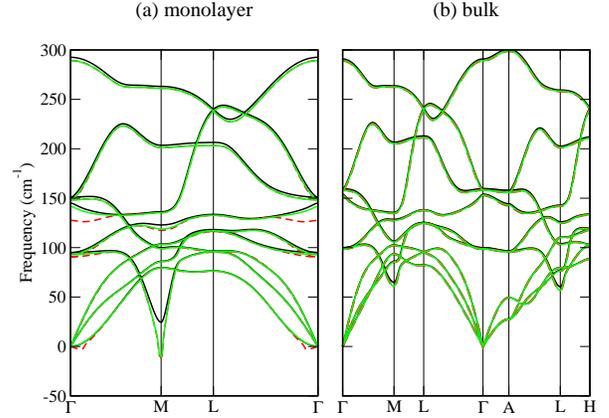}
    \caption{The individual contribution in SSCHA beyond the harmonic approximation at 0K. Black, red and green curves label the  phonon frequencies obtained from $\od$, $\od+\bDcal^{\bubble}$ and $\od+\bDcal^{\bubble}+\bDcal^{other}$ , respectively.}
    \label{fig:sscha-D3V-D4V-momo-bulk-0K}
\end{figure}

\begin{figure}
    \centering
    \includegraphics[width=0.9\linewidth]{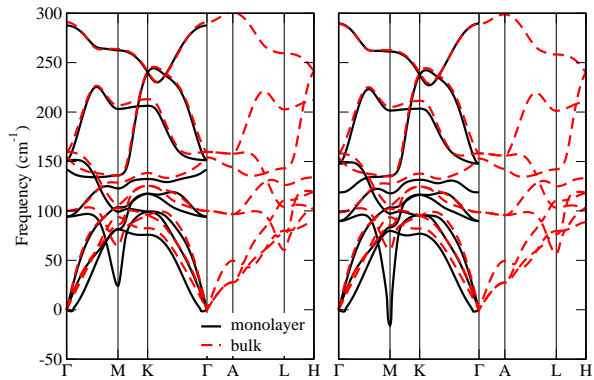}
    \caption{The phonon frequencies calculated using $\od$ (left panel) and $\od+\bDcal^{\bubble}$ (right panel) for monolayer and bulk \tite\ at 0K.}
    \label{fig:analysis-mat2R-mat3R-mono-bulk}
\end{figure}

Here we provide a detailed analysis of all the different anharmonic terms contributing to the free energy Hessian.
Within the SSCHA, the temperature dependent phonons are obtained from the dynamical matrix
\begin{equation}
\bDcal=\left.{\bf M}^{-\frac{1}{2}}\frac{\partial^2{F}}{\partial\bRcal \partial\bRcal}\right|_{\bRcal_{eq}} {\bf M}^{-\frac{1}{2}}
\end{equation}
where ${\bf M}$ is the matrix of the ionic masses $M_a$ with $M_{ab}=\delta_{ab}M_a$, and $\eval{\pdv{F}{\bRcal}{\bRcal}}_{\bRcal_{eq}}$ is the free energy Hessian with respect to the centroid positions ${\bf \Rcal}$ reads~\cite{Raffaello-PRB-2017}:
\begin{equation}
\label{eq:free-energy-hessian}
   \pdv{F}{\bRcal}{\bRcal}=\bPhi +\overset{(3)}{\bPhi}\bLambda(0) \overset{(3)}{\bPhi}+\overset{(3)}{\bPhi}\bLambda(0){\bf \Theta} \bLambda(0)\overset{(3)}{\bPhi} \, ,
\end{equation}
where $\bPhi$ represents the SSCHA force constant, $\overset{(3)}{\bPhi}\bLambda(0)\overset{(3)}{\bPhi}$ is the so-called ``static bubble term'', and $\overset{(3)}{\bPhi}\bLambda(0){\bf \Theta} \bLambda(0)\overset{(3)}{\bPhi}$ contains the higher order terms. Here $\overset{(n)}{\bPhi}$ refers to the $n$-th order anharmonic force constants averaged over the density matrix of the SSCHA hamiltonian 
(see Ref.~\cite{Raffaello-PRB-2017} for more details on notation). All these quantities can be recasted as appropriate stochastic averages over the atomic forces. 
The corresponding dynamical matrix can be written as:
\begin{equation}
\label{eq:dynexpansion}
   \bDcal= \od+\bDcal^{\bubble}+\bDcal^{other} \, ,
\end{equation}
where
\begin{subequations}
\begin{eqnarray}
\od&=&{\bf M}^{-\frac{1}{2}}\,\,\bPhi \,\, {\bf M}^{-\frac{1}{2}} \label{eq:od2} \, ;\\
\bDcal^{\bubble}&=& {\bf M}^{-\frac{1}{2}} \,\, \overset{(3)}{\bPhi}\bLambda(0) \overset{(3)}{\bPhi} \,\, {\bf M}^{-\frac{1}{2}} \label{eq:od3}\, ; \\
\bDcal^{other}&=& \bf {M}^{-\frac{1}{2}} \,\, \overset{(3)}{\bPhi}\bLambda(0){\bf \Theta} \bLambda(0)\overset{(3)}{\bPhi}\,\,  \bf {M}^{-\frac{1}{2}} \label{eq:od4} \,.
\end{eqnarray}
\end{subequations}
Analogously, with $\odo$ we refer to the harmonic dynamical matrix: 
\begin{equation}
 \odo=\left.{\bf M}^{-\frac{1}{2}}\,\,\frac{\partial^2 V}{\partial \bR \partial \bR}\right|_{\bR_{0}} \,\, {\bf M}^{-\frac{1}{2}}\, ,   
\end{equation}
where $\eval{\pdv{V}{\bR}{\bR}}_{\bR_{0}}$ is the Born-Oppenheimer potential energy Hessian in the `classical' equilibrium configuration $\bR_{0}$. The function $\bLambda(0)$ (See Eq.~(22) in Ref.~\cite{Raffaello-PRB-2017}) is mainly determined by the eigenvectors and eigenvalues of $\od$.
The different contributions to the dynamical matrix are shown in Fig. \ref{fig:sscha-D3V-D4V-momo-bulk-0K}. As it can be seen, the 
contributions arising from $\bDcal^{\bubble}$ and $\bDcal^{other}$ are negligible for the bulk case. In the monolayer, $\bDcal^{other}$ is negligible, 
however $\bDcal^{\bubble}$ is non negligible and it is the term responsible for the occurrence of the charge density wave.
A clearer comparison between the $\od$ and $\bDcal^{\bubble}$ in bulk and monolayer is shown in Fig. \ref{fig:analysis-mat2R-mat3R-mono-bulk}, underlining again the role of $\bDcal^{\bubble}$ in the monolayer.

\bibliography{main}

\end{document}